\documentclass[runningheads]{llncs}

\usepackage[T1]{fontenc}
\usepackage[utf8]{inputenc}
\usepackage[american]{babel}
\usepackage[all]{nowidow}
\usepackage[hyphens]{url}
\usepackage{amsmath}
\usepackage{amssymb}
\usepackage{amstext}
\usepackage{amsfonts}
\usepackage{array}
\usepackage{booktabs}
\usepackage{csquotes}
\usepackage[shortlabels]{enumitem}
\usepackage{graphicx}
\usepackage{multirow}

\begin{document}

\author{Laurent Chuat \and Sarah Plocher \and Adrian Perrig}
\institute{ETH Zurich}

\title{Zero-Knowledge User Authentication:\\ An Old Idea Whose Time Has Come}

\maketitle

\begin{abstract}
User authentication can rely on various factors (e.g., a password, a cryptographic key, and/or
biometric data) but should not reveal any secret information held by the user. This seemingly
paradoxical feat can be achieved through zero-knowledge proofs. Unfortunately, naive password-based
approaches still prevail on the web. Multi-factor authentication schemes address some of the
weaknesses of the traditional login process, but generally have deployability issues or degrade
usability even further as they assume users do not possess adequate hardware. This assumption no
longer holds: smartphones with biometric sensors, cameras, short-range communication capabilities,
and unlimited data plans have become ubiquitous. In this paper, we show that, assuming the user has
such a device, both security and usability can be drastically improved using an augmented
password-authenticated key agreement (PAKE) protocol and message authentication codes.
\end{abstract}

\section{Introduction}

User authentication still typically involves transmitting a plaintext password to the web server,
which is problematic for many reasons. First, an attacker could obtain the user's password using
malware, a keystroke logger, phishing, or a man-in-the-middle attack. Second, the user might be
using the same password on different websites, allowing a malicious server operator to impersonate
the user. Third, if the website's database is breached, although passwords may be protected by a
hash function, a brute-force attack could reveal low-entropy passwords. Fourth, as password-reuse
and low-entropy passwords are problematic, users are constantly advised to pick new, long,
complicated passwords. Finally, transmitting the password over a secure channel, storing it as a
hash, and using a random salt as well as a strong cryptographic hash function are recommended
measures, but they are impossible to enforce or even verify for the end user. The unsalted
SHA-1-hashed passwords of millions of LinkedIn users were disclosed in 2012~\cite{kamp2012linkedin}
and hundreds of millions more were disclosed (with corresponding email addresses) in
2016~\cite{linkedin2016}, for example.

Two-factor authentication (2FA) supposedly reinforces standard passwords. Typically on today's web,
the time-based one-time password (TOTP) algorithm~\cite{rfc6238} is used to generate a new 6-digit
code every 30 seconds, based on the current time and a secret shared between the server and the
user's authentication device (i.e., either a dedicated token or a smartphone). Usually, the user
must submit their one-time code after the traditional username/password step. For this reason,
two-factor authentication is also referred to as ``2-step verification''~\cite{google_auth}.

Although 2FA may protect users against certain attacks, it doesn't completely prevent them.
Schneier~\cite{schneier2005cacm} claims that it ``doesn't solve anything''. Indeed, an attacker
could still obtain the password and a one-time code---or hijack the session---through a real-time
attack~\cite{real_time}. The attacker could then ``make any fraudulent transaction he wants'', which
includes changing the password and deactivating 2FA to gain long-term control over the user's online
account. Passwords are still weak and typed on untrusted devices. Servers still receive plaintext
passwords. Password re-use is still problematic. All secret values may still be known by the server.
Finally, assuming that an attacker manages to uncover a large number of passwords, the entropy of a
6-digit code is too low to rule out a brute-force attack. Therefore, TOTP-based authentication
alleviates neither website owners nor users from the consequences of password leaks.

Herein we challenge some of the assumptions underlying existing schemes. For example, should we
still assume that authentication devices have no Internet access? Clearly, the fact that a
smartphone may not have Internet access at all times should be taken into account, but such a
situation has become the exception rather than the rule. Also, is it sufficient to supplement a weak
form of authentication with a ``second step''? We believe, on the contrary, that user authentication
should be completely reassessed.

Besides security, usability is actually deteriorated by schemes based on the TOTP algorithm. Every
time users want to log in, they must type an ever-changing code displayed on their device---in
addition to their password. The scheme we present, ZeroTwo, on the other hand, would fit inside a
streamlined process, where the user only enters a username or email address on the website and then
approves the login/authorization request in a single step on their smartphone. ZeroTwo relies on an
augmented PAKE\footnote{also referred to as \emph{asymmetric} password-authenticated key
establishment or aPAKE} protocol and message authentication codes (MACs) to provide the server with
a zero-knowledge password proof (ZKPP) and evidence that the legitimate user explicitly authorized
critical actions.

Another problem that has been largely overlooked in previous work is that users share their
credentials with friends and family~\cite{singh2007password}. This problem can be solved by moving
critical parts of the authentication process to the user's smartphone, which allows them to remotely
give someone they trust a restricted, temporary, and revokable access to their online account.

\section{Protocol Overview}

In this section, we provide an overview of how users can authenticate and explicitly authorize
actions using ZeroTwo. We chose to base our scheme upon SRP~\cite{wu2002srp} because mature
implementations are available. We also borrowed concepts from AugPAKE~\cite{rfc6628}. In principle,
however, any PAKE protocol (e.g., OPAQUE~\cite{jarecki2018opaque}) could be used to develop a scheme
similar to ours.

\subsection{Notation}

Table~\ref{fig:notation} shows the notation we use throughout the paper. The values $n$ and $g$ are
agreed upon beforehand. All arithmetic is performed modulo $n$. $\text{H}()$ and $\text{H}_K()$
denote a hash function and an HMAC, respectively.

\begin{table}[h!]
    \centering
    \caption{Notation Summary}
    \begin{tabular}{r @{\hskip .5em} p{185pt}}
        \toprule
        $n$ & large safe prime number, i.e., $n = 2q + 1$ where $q$ is also prime \\
        $g$ & primitive root modulo $n$ \\
        $k$ & multiplier parameter, $k =$ H($n$, $g$) \\
        $l$ & public parameter, $l =$ (H($n$) xor H($g$)) \\
        $I_u$ & user's identifier (username or email address) \\
        $I_s$ & server's identifier (domain name) \\
        $P$ & password and/or biometric input \\
        $p$ & master secret (see Section~\ref{sec:secret_nature}) \\
        $x$ & effective secret, discarded after the computation of the verifier, $x =
            \text{H}(I_u, I_s, p)$ \\
        $v$ & verifier, $v = g^x$ \\
        $a, b$ & ephemeral private keys \\
        $A, B$ & corresponding public keys \\
        $u$ & scrambling parameter, $u =$ H($A$, $B$) \\
        $S$ & common exponential value \\
        $K$ & session key \\
        $M$ & authorization (evidence that $K$ is known) \\
        $d$ & duration of the session \\
        $o$ & operation to authorize \\
        \bottomrule
    \end{tabular}
    \label{fig:notation}
\end{table}

\subsection{Initialization}

During sign up, the following operations must be performed:

\begin{enumerate}
    \item The user chooses a unique identifier $I_u$ and sends it to the server (from a browser).
    \item The server replies with a QR code containing both the user identity $I_u$ and the server
    identity $I_s$ as well as the URL that the smartphone must use to send the initialization data.
    \item The user scans the QR code; chooses a master secret $p$ (see
    Section~\ref{sec:secret_nature}), if this has not been done before; and the app computes the
    following:
    \begin{equation*}
    	x = \text{H}(I_u, I_s, p)\footnote{As in AugPAKE~\cite{rfc6628} and other previous
    	work~\cite{bonneau2011getting}, but unlike in SRP~\cite{wu2002srp}, using a salt is
    	unnecessary here, as the effective secret is derived from unique identifiers.}
    \end{equation*}
    \begin{equation*}
    	v = g^x
    \end{equation*}
    \item The user's identity and the verifier are sent to the previously received URL.
    \item The server stores the verifier permanently (if the identifier is valid and not already
    used) under the user's identifier. If the identifier is an email address, the server should
    first verify that the user has access to that address.
\end{enumerate}

\subsection{Authentication}

The authentication process works as follows:

\begin{enumerate}
    \item The user initiates a login attempt by sending their identifier $I_u$ to the server (from a
    browser).
    \item The server looks up the user's verifier $v$, generates a random ephemeral private key $b$,
    and computes the corresponding public key:
    \begin{equation*}
    	B = kv + g^b
    \end{equation*}
    \item The server sends its public key $B$ as well as identifiers $I_u$ and $I_s$ to both the
    browser and smartphone. In the common case, the smartphone is connected to the Internet and
    receives $B$ (with a notification), a fingerprint~\cite{dechand2016empirical,tan2017can} of the
    public key is displayed on both the smartphone app and the browser, and the user is asked to
    make sure that the two fingerprints match. If the smartphone is not connected to the Internet,
    the user must select a method for transferring $B$ from the browser to the app (see
    Section~\ref{sec:alternative}).
    \item The user can accept the authentication request on their smartphone by entering a password
    $P$ and/or using an embedded biometric sensor. The master secret $p$ is then used to compute the
    session key $K$, and $M$, which constitutes evidence that the user authorized a session to be
    established for a time period $d$:
    \begin{equation*}
    	x = \text{H}(I_u, I_s, p)
    \end{equation*}
    \begin{equation*}
    	a = \text{random}()
    \end{equation*}
    \begin{equation*}
    	A = g^a
    \end{equation*}
    \begin{equation*}
    	u = H(A, B)
    \end{equation*}
    \begin{equation*}
    	S = (B - kg^x)^{(a + ux)}
    \end{equation*}
    \begin{equation*}
    	K = \text{H}(S)
    \end{equation*}
    \begin{equation*}
    	M = \text{H}_K(l, I_u, I_s, A, B, d)
    \end{equation*}
    \item The user's identifier $I_u$, public key $A$, and proof $M$, as well as the authorized
    duration of the session $d$ are transmitted to the server by the smartphone.
    \item The server verifies $M$ by computing its own session key:
    \begin{equation*}
    	S = (Av^u)^b
    \end{equation*}
    \begin{equation*}
    	K = \text{H}(S)
    \end{equation*}
    If the received data is correct, the server automatically redirects the client to the page that
    required authentication.
\end{enumerate}

\subsection{Explicit Authorization}

Following the principle of least privilege and to mitigate session hijacking, web developers using
ZeroTwo may choose to define a set of actions for which explicit authorization from the user is
required. When such an action is requested by the client, the server generates a human-readable
message $o$ describing the action and sends it directly to the smartphone, with a notification and a
random nonce $c$ (for replay prevention). The smartphone then computes an authorization message $M$
with the session key $K$:
\begin{equation*}
M = \text{H}_K(o, c)
\end{equation*}
and sends it to the server. This can also be performed through alternative channels, as described in
Section~\ref{sec:alternative}.

This general concept is sometimes referred to as ``transaction signing''. In the context of this
paper, however, because a shared key is established between the client and the server, symmetric
cryptography is sufficient.

\subsection{Session Management}

The session key $K$ is stored on the smartphone and considered valid by the server for as long as
the session is valid (specified with $d$).

If the user wants to terminate the session before it expires automatically, then they can send an
authenticated logout message $M$ to the server:
\begin{equation*}
M = \text{H}_K(o), \quad\text{where }o = \texttt{logout}.
\end{equation*}

The user should also be able to logout in the usual way, i.e., from the web browser, but the above
method allows the user (who forgot to logout or was prevented from doing so by an attacker, for
example) to terminate the session from their smartphone.

\subsection{Alternative Channels}
\label{sec:alternative}

A bidirectional communication channel between the smartphone and the server is needed during the
authentication and authorization procedures. The server's public key $B$ can simply be displayed as
a QR code in the browser and scanned with the ZeroTwo app. Sending data back to the server from the
smartphone is more challenging, because it may not be possible to establish a direct Internet
connection. The WebRTC standard, through a collection of protocols and APIs, allows modern web
browsers to access peripheral hardware, which provides several solutions to our problem:
\begin{itemize}
\item \textbf{Webcam:} A camera connected to or embedded in the terminal (e.g.,
laptop) is used to scan a QR code displayed on the smartphone.
\item \textbf{Bluetooth:} A Bluetooth connection is established between the
smartphone and the browser to transfer the authentication data.
\end{itemize}
The user may initiate the authentication process directly from a mobile browser. In that case, the
above methods cannot be used. By definition, however, the smartphone would need an Internet
connection to start the authentication procedure from the login page and thus an alternative channel
would not be needed.

\section{Nature of the Master Secret}
\label{sec:secret_nature}

One of the advantages of augmented PAKE protocols is that they provide a great deal of flexibility
with regard to the nature of the secret they rely upon. In other words, the master secret (which we
denote $p$), although it is traditionally assumed to be a password, can be arbitrary long and of any
nature. And because of the zero-knowledge property of our protocol, no entity can learn anything
about the secret aside from the one who created it. Therefore, the responsibility of developing a
process for generating the master secret that is both secure and convenient for users lies entirely
with the developers of compatible apps.

Our preferred approach consists in generating a random passphrase (e.g., composed of lowercase,
hyphen-separated words). A passphrase has the advantage that it can be memorized and/or written on a
piece of paper to be kept in a secure location. Moreover, a long passphrase generated at random from
a large set of words has enough entropy to make brute-force and dictionary attacks infeasible. The
passphrase can be re-used for several websites because the effective secret $x$ is domain-dependent
and the entropy of the master secret is high. The only disadvantage of a passphrase would be to type
it at every login, but it can be stored on the smartphone.

One could also combine several secrets. Concatenating a key and a password, for example, would
result in a multi-factor scheme. If the same combination is used on multiple websites, however,
changing the password is inconvenient as all verifiers must be updated. Therefore, it is not clear
whether the presumed security gain would be worth the decrease in usability. Instead, we propose to
protect a high-entropy secret with another factor, as described in Section~\ref{sec:storage}.

A single high-entropy secret is sufficient to protect multiple websites. However, to avoid a single
point of failure, some users might choose to generate multiple secrets: one for personal websites
and one for professional purposes, for example. Therefore, compatible apps should offer an option to
manage multiple accounts.

\section{Access Control, Storage, and Backup}
\label{sec:storage}

If the master secret is stored on the smartphone, then the access to the ZeroTwo-compatible app and
the storage of the secret should be protected. The two main options at our disposal are the
following:

\begin{itemize}
    \item \textbf{Biometric protection:} Fingerprint and facial scanners are becoming commonplace on
    high-end smartphones. Android and iOS both provide APIs for letting developers protect their app
    with embedded biometric sensors.  
    \item \textbf{Password protection:} Although passwords do not allow for the same ease of use as
    biometrics, they offer another advantage: the master secret can be encrypted before a backup
    (in cloud storage, for example).
\end{itemize}

Although we favor the biometrics approach, a password can be used as a fallback in case the
smartphone does not have a biometric sensor, or in case the biometric sensor produces a false
negative. As for the backup strategy, assuming the master secret is a passphrase, users should be
able to decide whether they want to only display the passphrase when it is first generated (to
memorize it or write it down) or back it up using a protected cloud storage solution.

We note that a single-secret approach, as opposed to a naive password-based scheme, scales to an
arbitrary number of websites without having to constantly remember/backup new secrets, which is
particularly important should the user decide to use an offline backup solution.

\section{Comparative Evaluation and Related Work}

The last couple of decades has seen a plethora of proposals for user authentication. In general,
existing schemes suffer from at least one of the following drawbacks: (a) they require a dedicated
device, (b) they are proprietary, (c) they involve a shared secret, and/or (d) they still require a
traditional password. Herein we only discuss a small subset of schemes and refer to the paper by
Bonneau et al.~\cite{JosephBonneau:2012un} for an extensive evaluation of related work. Using their
framework we evaluated ZeroTwo and present the results in Table~\ref{tab:comparison}.

\begin{table}[h!]
	\centering
	\renewcommand{\arraystretch}{1.1}
	\newcommand*\rot{\rotatebox{90}}
	\newcommand{\nope}{\Large~}
	\newcommand{\crcl}{\Large$\circ$}
	\newcommand{\bull}{\Large$\bullet$}
	\begin{tabular}{@{} l |
		@{\enskip} *{8}{@{}>{\centering}b{1.2em}@{}} @{\enskip} |
		@{\enskip} *{6}{@{}>{\centering}b{1.2em}@{}} @{\enskip} |
		@{\enskip} *{11}{@{}>{\centering}b{1.2em}@{}}
		}
        & \multicolumn{8}{c | @{\enskip}}{\textbf{Usability}}
        & \multicolumn{6}{c | @{\enskip}}{\hspace{-4pt}\textbf{Deployability}}
        & \multicolumn{11}{c}{\textbf{Security}} \\[2ex]
        \textbf{Scheme} &
        	\rot{Memory-wise effortless} &
        	\rot{Scalable for users} &
        	\rot{Nothing to carry} &
        	\rot{Physically effortless} &
        	\rot{Easy to learn} &
        	\rot{Efficient to use} &
        	\rot{Infrequent errors} &
        	\rot{Easy recovery from loss} &
        	\rot{Accessible} &
        	\rot{Negligible cost per user} &
        	\rot{Server-compatible} &
        	\rot{Browser-compatible} &
        	\rot{Mature} &
        	\rot{Non-proprietary} &
        	\rot{Resilient to physical observation} &
        	\rot{Resilient to targeted impersonation} &
        	\rot{Resilient to throttled guessing} &
        	\rot{Resilient to unthrottled guessing} &
        	\rot{Resilient to internal observations} &
        	\rot{Resilient to leaks from other verifiers} &
        	\rot{Resilient to phishing} &
        	\rot{Resilient to theft} &
        	\rot{No trusted third party} &
        	\rot{Requiring explicit consent} &
        	\rot{Unlinkable}
        	\tabularnewline
        \hline
        Passwords   & \nope & \nope & \bull & \nope & \bull & \bull & \crcl & \bull & \bull & \bull & \bull & \bull & \bull & \bull & \nope & \crcl & \nope & \nope & \nope & \nope & \nope & \bull & \bull & \bull & \nope \tabularnewline
        \hline
        ZeroTwo$^1$ & \bull & \bull & \crcl & \crcl & \bull & \bull & \crcl & \bull & \bull & \crcl & \nope & \bull & \nope & \bull & \bull & \bull & \bull & \bull & \crcl & \bull & \bull & \bull & \bull & \bull & \bull \tabularnewline
        ZeroTwo$^2$ & \nope & \bull & \crcl & \nope & \bull & \crcl & \crcl & \bull & \crcl & \crcl & \nope & \bull & \nope & \bull & \bull & \bull & \bull & \bull & \crcl & \bull & \bull & \bull & \bull & \bull & \bull \tabularnewline
        \hline
    \end{tabular} \\[2ex]
    
    \normalsize
    $\bullet = $ offers the benefit; $\circ = $ almost offers the benefit
    \smallskip
    
	\caption{Comparative evaluation of ZeroTwo, assuming a single passphrase is used as the master
	secret. $^1$\emph{Best-case scenario:} The smartphone has a biometric sensor to protect the
	secret and Internet connectivity is available. $^2$\emph{Worst-case scenario:} The passphrase is
	protected with a password; the smartphone has no biometric sensor and no Internet connectivity.}
	\label{tab:comparison}
\end{table}

Bonneau~\cite{bonneau2011getting} had previously proposed a password-based authentication protocol,
designed to avoid revealing the password to the server (using Javascript), which requires neither
a software update on the client side nor a separate authentication device. 
Thomas et al.~\cite{thomas2014better} similarly focused on restrictions imposed by legacy systems to
address the issues of weak passwords.

Hardware tokens (such as YubiKey~\cite{yubikey} or Pico~\cite{pico}) have the advantage that they
are shielded from remote attackers. Although they are particularly adapted to
professional contexts, we believe that they are less suited to the general public as many users are
unwilling to spend money for a dedicated device.

Cronto~\cite{cronto} is a phone-based scheme close to our proposal in terms of offered features, but
it is a proprietary product whose exact design is not publicly known, which makes it hard to analyze
and unlikely to be widely deployed outside of the banking industry.
SQRL (pronounced ``squirrel'') or Secure Quick Reliable Login~\cite{SQRL} also offers a strong
phone-based authentication solution, but it does not support transaction signing or explicit
authorization.

Sound-Proof~\cite{sound_proof} is a recent system that relies on sound for the smartphone and
browser to communicate. One of the main goals of Sound-Proof is to provide a seamless experience to
users, i.e., the phone need not even be handled for the authentication process to complete. However,
the user still has to type a password in the browser, which comes with the issues we discussed
previously. Moreover, the complete seamlessness of Sound-Proof is not compatible with our view that
certain actions should be explicitly authorized on a trusted device.

The FIDO Alliance has published a number of protocol, framework, and API specifications~\cite{FIDO}
for user authentication on the web, based on public-key cryptography. To the best our knowledge,
none of these specifications rely on password-authenticated key establishment.

\section{Conclusion}

Many forms of user authentication on today's web suffer from severe drawbacks. Using public-key
cryptography for authentication is not a novel idea. But whereas asymmetric cryptography is commonly
used for server authentication on the web (through TLS), it is not a widespread approach for user
authentication. There are several reasons for this: key management is difficult; users need a
trusted device with storage, communication, and computation capabilities; and the authentication
process must be fast and convenient. We believe, however, that the democratization of smartphones
with embedded biometric sensors, unlimited cellular data plans, and new communication standards such
as WebRTC now make asymmetric protocols a viable option for user authentication.

\section*{Acknowledgments}

We gratefully thank Eduardo Solana for his valuable input in the early stages of this project,
Daniel R. Thomas for his extensive feedback, and all the workshop attendees who participated in the
discussion and helped improve this paper.

\bibliographystyle{plain}
\bibliography{bibliography}

\begin{thebibliography}{10}

\bibitem{bonneau2011getting}
Joseph Bonneau.
\newblock Getting web authentication right: A best-case protocol for the
  remaining life of passwords.
\newblock In {\em Proceedings of the 19th International Workshop on Security
  Protocols}, 2011.

\bibitem{JosephBonneau:2012un}
Joseph Bonneau, Cormac Herley, Paul~C. van Oorschot, and Frank Stajano.
\newblock The quest to replace passwords: A framework for comparative
  evaluation of web authentication schemes.
\newblock In {\em Proceedings of the 33rd IEEE Symposium on Security and
  Privacy (S\&P)}, 2012.

\bibitem{SQRL}
Gibson~Research Corporation.
\newblock {SQRL} secure quick reliable login.
\newblock \url{https://www.grc.com/sqrl/sqrl.htm}.

\bibitem{dechand2016empirical}
Sergej Dechand, Dominik Schürmann, Karoline Busse, Yasemin Acar, Sascha Fahl,
  and Matthew Smith.
\newblock An empirical study of textual key-fingerprint representations.
\newblock In {\em Proceedings of the 25th USENIX Security Symposium}, 2016.

\bibitem{linkedin2016}
Lorenzo Franceschi-Bicchierai.
\newblock Another day, another hack: 117 million {LinkedIn} emails and
  passwords.
\newblock \url{https://perma.cc/6MC6-EVHH}, May 2016.
\newblock Motherboard.

\bibitem{google_auth}
Google.
\newblock 2-step verification.
\newblock \url{https://www.google.com/landing/2step}.

\bibitem{jarecki2018opaque}
Stanislaw Jarecki, Hugo Krawczyk, and Jiayu Xu.
\newblock {OPAQUE}: An asymmetric {PAKE} protocol secure against
  pre-computation attacks.
\newblock In {\em Proceedings of the 37th Annual International Conference on
  the Theory and Applications of Cryptographic Techniques (Eurocrypt)}, 2018.

\bibitem{kamp2012linkedin}
Poul-Henning Kamp.
\newblock {LinkedIn} password leak: Salt their hide.
\newblock {\em ACM Queue}, 10(6):20, June 2012.

\bibitem{sound_proof}
Nikolaos Karapanos, Claudio Marforio, Claudio Soriente, and Srdjan Capkun.
\newblock {Sound-Proof}: Usable two-factor authentication based on ambient
  sound.
\newblock In {\em Proceedings of the 24th USENIX Security Symposium}, 2015.

\bibitem{rfc6238}
David M'Raihi, Salah Machani, Mingliang Pei, and Johan Rydell.
\newblock {TOTP}: Time-based one-time password algorithm.
\newblock RFC 6238, May 2011.

\bibitem{cronto}
OneSpan.
\newblock {CRONTO} mobile app.
\newblock \url{https://perma.cc/THZ6-3YFW}.

\bibitem{schneier2005cacm}
Bruce Schneier.
\newblock Two-factor authentication: too little, too late.
\newblock {\em Communications of the ACM}, 48(4):136, April 2005.

\bibitem{real_time}
Bruce Schneier.
\newblock Real-time attacks against two-factor authentication.
\newblock \url{https://perma.cc/FQ9R-USG6}, December 2018.
\newblock Schneier on Security.

\bibitem{rfc6628}
SeongHan Shin and Kazukuni Kobara.
\newblock Efficient augmented password-only authentication and key exchange for
  {IKEv2}.
\newblock RFC 6628, June 2012.

\bibitem{singh2007password}
Supriya Singh, Anuja Cabraal, Catherine Demosthenous, Gunela Astbrink, and
  Michele Furlong.
\newblock Password sharing: Implications for security design based on social
  practice.
\newblock In {\em Proceedings of the SIGCHI Conference on Human Factors in
  Computing Systems}, 2007.

\bibitem{pico}
Frank Stajano.
\newblock Pico: No more passwords!
\newblock In {\em Proceedings of the 19th International Workshop on Security
  Protocols}, 2011.

\bibitem{tan2017can}
Joshua Tan, Lujo Bauer, Joseph Bonneau, Lorrie~Faith Cranor, Jeremy Thomas, and
  Blase Ur.
\newblock Can unicorns help users compare crypto key fingerprints?
\newblock In {\em Proceedings of the SIGCHI Conference on Human Factors in
  Computing Systems}, 2017.

\bibitem{FIDO}
{The FIDO Alliance}.
\newblock Specifications overview ({FIDO2}, {WebAuthn}, {FIDO UAF}, {FIDO
  U2F}).
\newblock \url{https://fidoalliance.org/specifications}.

\bibitem{thomas2014better}
Daniel~R. Thomas and Alastair~R. Beresford.
\newblock Better authentication: Password revolution by evolution.
\newblock In {\em Proceedings of the 22th International Workshop on Security
  Protocols}, 2014.

\bibitem{wu2002srp}
Thomas Wu.
\newblock {SRP-6}: Improvements and refinements to the {Secure Remote Password}
  protocol.
\newblock {IEEE P1363 Working Group}, October 2002.

\bibitem{yubikey}
Yubico.
\newblock {YubiKey} strong two factor authentication for business and
  individual use.
\newblock \url{https://www.yubico.com}.

\end{thebibliography}

\end{document}